%% file: GlobalSIP_Arxive.tex
\documentclass[conference]{IEEEtran}
\usepackage{fixltx2e}
\usepackage{amsmath, amsthm, amssymb, bm}
\usepackage{epsfig}
\usepackage{amsfonts}
\usepackage{epstopdf}
\usepackage{subcaption}
\usepackage{cite}

\usepackage{tikz,pgfplots} 

\newlength\figureheight
\newlength\figurewidth
\usepackage[autostyle]{csquotes}

\include{macros}

\begin{document}

\title{Robust Regularized Least-Squares Beamforming Approach to Signal Estimation}

%\author{\IEEEauthorblockN{Mohamed~Suliman,
%Tarig~Ballal, Abla~Kammoun, and Tareq~Y.~Al-Naffouri.}
%\\ \IEEEauthorblockA{  
%\vspace{7pt}
% Computer, Electrical, and Mathematical Sciences and Engineering (CEMSE) Division,\\
%King Abdullah University of Science and Technology (KAUST),
%Thuwal, Makkah Province, Saudi Arabia. \\ Emails: $\{$mohamed.suliman, tarig.ahmed, abla.kammoun, tareq.alnaffouri$\}$@kaust.edu.sa 
%} 
%}

\author{ Mohamed~Suliman\thanks{\textcopyright 2016 IEEE. Personal use of this material is permitted. Permission from IEEE must be obtained for all other uses, in any current or future media, including reprinting/republishing this material for advertising or promotional purposes, creating new collective works, for resale or redistribution to servers or lists, or reuse of any copyrighted component of this work in other works.

This work was funded in part by a CRG2 grant CRG\_R2\_13\_ALOU\_KAUST\_2 from the Office of Competitive Research (OCRF) at King Abdullah University of Science and Technology (KAUST). }, Tarig~Ballal, and Tareq~Y.~Al-Naffouri.  
\\ Computer, Electrical, and Mathematical Sciences and Engineering (CEMSE) Division,\\
 King Abdullah University of Science and Technology (KAUST), Thuwal, Makkah Province, Saudi Arabia.\\
 Emails: $\{$mohamed.suliman, tarig.ahmed, tareq.alnaffouri$\}$@kaust.edu.sa 
}
%\author{ Mohamed Suliman, Tarig Ballal, and Tareq Y. Al-Naffouri,~\IEEEmembership{Member, IEEE}% <-this % stops a space
%%\begin{small}
%%\\{Computer, Electrical, and Mathematical Sciences and Engineering (CEMSE) Division,\\
%%King Abdullah University of Science and Technology (KAUST),\\
%%Thuwal, Makkah Province, Saudi Arabia.}
%%\end{small}% <-this % stops a space
%
%\thanks{M. Suliman, T. Ballal, and T. Y. Al-Naffouri are with the Electrical Engineering Department, King Abdullah University of Science and Technology (KAUST), Thuwal, Makkah Province, Saudi Arabia. E-mails:$\{$mohamed.suliman, tarig.ahmed, tareq.alnaffouri$\}$@kaust.edu.sa.}
%
%}

%\author{    \fontsize{9.3pt}{9.3pt} \vspace{3.4pt} Mohamed Suliman, Tarig~Ballal, Abla~Kammoun, and
%Tareq~Y.~Al-Naffouri\\ 
%Computer, Electrical, and Mathematical Sciences and Engineering (CEMSE) Division,\\
% King Abdullah University of Science and Technology (KAUST),
%Thuwal, Makkah Province, Saudi Arabia.\\ Emails: $\{$mohamed.suliman, tarig.ahmed, abla.kammoun, tareq.alnaffouri$\}$@kaust.edu.sa 
%}

\IEEEoverridecommandlockouts

\maketitle

\begin{abstract}
In this paper, we address the problem of robust adaptive beamforming of signals received by a linear array. The challenge associated with the beamforming problem is twofold. Firstly, the process requires the inversion of the usually ill-conditioned covariance matrix of the received signals. Secondly, the steering vector pertaining to the direction of arrival of the signal of interest is not known precisely. To tackle these two challenges, the standard capon beamformer is manipulated to a form where the beamformer output is obtained as a scaled version of the inner product of two vectors. The two vectors are linearly related to the steering vector and the received signal snapshot, respectively. The linear operator, in both cases, is the square root of the covariance matrix. A \emph{regularized least-squares} (RLS) approach is proposed to estimate these two vectors and to provide robustness without exploiting prior information. Simulation results show that the RLS beamformer using the proposed regularization algorithm outperforms state-of-the-art beamforming algorithms, as well as another RLS beamformers using a standard regularization approaches.
\end{abstract}
 
%%%%%%%%%%%%%%%%%%%%%%%%%%%%%%%%%%%%%%%%%%%%%%%%%%%%%%%%%%%%%%%%%%%%%%%%%%%%%%%%%%%%%%%%%%%%%%%%%%%%%%%%%%%%%%%%%%%%%%%%%%%%%%%%%%%%%%%%%%%%%%%%%%%%%%%%%%%%%%%%%%%%%%%%%%%%%%%%%%%%%%%%%%%%%%%%%%%%%%%%%%%%%%%%%%%%%%%  
\section{Introduction}
\label{sec:intro}
Robust adaptive beamforming is concerned with the alleviation of the impact of two problems. The first problem is related to the covariance matrix of the received signals. This matrix is typically estimated using a finite number of signal snapshots, which results in an ill-conditioned matrix. It is well known that the inversion of ill-conditioned matrices as part of a problem solution leads to inaccuracy in that solution. A simple solution to this problem is by diagonal loading of the sample covariance matrix \cite{li2003robust}. The other problem in robust beamforming is the mismatch between the actual steering vector and the presumed steering vector. Such a mismatch is due to one of several causes that include environmental non-stationarities, look direction errors, imperfect array calibration, and distorted antenna shape, to mention a few \cite{Vorobyov2003}.
 
Throughout the years, many algorithms have been proposed for robust adaptive beamforming. A popular approach is the so-called \emph{linearly constrained minimum variance} (LCMV) beamformer \cite{Frost1972, Takao1976} which offers robustness against signal steering vector mismatch by imposing several linear constraints. In \cite{Xu2015}, a robust LCMV beamformer is devised by formulating the problem as a non-convex quadratically constrained quadratic programming problem.
 
Capon beamformer, also known as \emph{minimum variance distortionless} (MVDR) beamformer is notorious of being sensitive to the errors in the steering vector. The work in \cite{li2003robust} discusses the extension of Capon beamformer to the case where the steering vector is uncertain. In \cite{Vorobyov2003}, an approach based on optimizing the worst-case performance is proposed. This approach minimizes a quadratic function subject to infinitely many non-convex quadratic constraints. In \cite{Lorenz2005}, the uncertainty in the array manifold is explicitly modeled via an ellipsoid and the robust weight selection process is cast as a second-order cone program. Typically, the MVDR based robust beamformer relies on certain available prior information. The method proposed in \cite{Khabbazi2012} attempts to use as little prior information as possible. Mathematically, the proposed beamformer of \cite{Khabbazi2012} is expressed as a non-convex quadratically constrained quadratic programming problem.
 
In this work, we propose a new approach for robust beamforming based on the capon beamformer. We start from the MVDR solution. This solution is sensitive due to the effect of the ill-conditionedness of the signal covariance matrix and that of the uncertainty in the steering vector. We impose robustness by manipulating the MVDR solution to a form where the robust beamforming solution is equivalent to solving two linear systems. Then, a new regularization approach is proposed for solving these two linear systems and to provide the required robustness without using any prior information. Simulation results show that the RLS beamformer using the proposed regularization algorithm outperforms state-of-the-art beamforming algorithms.
%%%%%%%%%%%%%%%%%%%%%%%%%%%%%%%%%%%%%%%%%%%%%%%%%%%%%%%%%%%%%%%%%%%%%%%%%%%%%%%%%%%%%%%%%%%%%%%%%%%%%%%%%%%%%%%%%%%%%%%%%%%%%%%%%%%%%%%%%%%%%%%%%%%%%%%%%%%%%%%%%%%%%%%%%%%%%%%%%%%%%%%%%%%%%%%%%%%%%%%%%%%%%%%%%%%%%%%
\section{Regularized Least-squares (RLS)}
Let us consider the linear model
\begin{equation}
\label{eq:model}
\rv = \Am \xv + \vv,
\end{equation}
where $\Am \in \mathbb{C}^{m \times n}$ is a Hermitian matrix and $\vv$ is the noise vector that is assumed to be white Gaussian with unknown variance $\sigma_{\vv}^{2}$. A common approach to find an estimate of $\xv$ is by using the \emph{least-squares} (LS) method \cite{kay2013fundamentals} which is given by
\begin{eqnarray}
\label{eq:pure LS solution}
{\hat{\xv}}_{\text{LS}}
&=& (\Am^{H}  \Am)^{-1}  \Am^{H} \rv,
\end{eqnarray}
where $\left(.\right)^{H}$ is the Hermitian of the matrix. The difficulty with LS occurs when $\Am$ is \emph{ill-conditioned}. In such a case, (\ref{eq:pure LS solution}) is very sensitive to perturbations in the data. To overcome this difficulty, \emph{regularization methods} are frequently used \cite{tikhonov2013numerical}.

In this paper, we are particularly interested in the RLS estimator which is given by
\begin{equation}
\label{eq:RLS}
\hat{\xv}_{\text{RLS}} = (\Am^{H}  \Am + \gamma\Id)^{-1}  \Am^{H} \rv,
\end{equation}
where $\gamma \in \mathbb{R}^{+}$ is the \emph{regularization parameter} and $\Id$ is the identity matrix. Several regularization parameter selection methods have been proposed to find the regularization parameter. These methods include the \emph{L-curve} \cite{hansen1993use}, the \emph{generalized cross validation} (GCV) \cite{wahba1990spline}, and the \emph{quasi-optimal} method \cite{morozov2012methods, bauer2008regularization}, to name a few.
%%%%%%%%%%%%%%%%%%%%%%%%%%%%%%%%%%%%%%%%%%%%%%%%%%%%%%%%%%%%%%%%%%%%%%%%%%%%%%%%%%%%%%%%%%%%%%%%%%%%%%%%%%%%%%%%%%%%%%%%%%%%%%%%%%%%%%%%%%%%%%%%%%%%%%%%%%%%%%%%%%%%%%%%%%%%%%%%%%%%%%%%%%%%%%%%%%%%%%%%%%%%%%%%%%%%%%%
\section{The Proposed Beamforming Approach}
\label{sec:beam}
The output of a beamformer for an array with $n_{\text{e}}$ elements, at a discrete time instant $t$, is given by
\begin{align}
\label{eq:MVDR output}
y_{\text{BF}}[t]  =  \wv^H \yv[t],
\end{align}
where $\wv \in \mathbb{C}^{n_e}$ is the beamformer weighting coefficients  vector while $\yv[t] \in \mathbb{C}^{n_e}$ is the array observations \enquote{snapshots} vector. For the Capon/MVDR beamformer, the weighing coefficients are given by \cite{li2003robust}
\begin{align}
\label{eq:MVDR weights}
\wv_{\text{MVDR}} = \frac{ \hat{\Cm}_{\yv\yv}^{-1} \av}{\av^H  \hat{\Cm}_{\yv\yv}^{-1} \av} ,
\end{align}
where $\av$ is the array \emph{steering vector} and $\hat{\Cm}_{\yv\yv}$ is the sample covariance matrix of the received signals, which is estimated from $n_{s}$ \emph{snapshots} according to
\begin{align}
\label{eq:sample cov}
 \hat{\Cm}_{\yv\yv} = \frac{1}{n_s} \sum_{t=1}^{n_s} \yv[t] \yv[t]^H.
\end{align}
As discussed previously, the difficulty with the MVDR beamformer is due to the ill-conditionedness of the matrix $\hat{\Cm}_{\yv\yv}$ and the uncertainty in the steering vector $\av$. Based on \eqref{eq:MVDR output} and \eqref{eq:MVDR weights}, we can write the beamformer output in (\ref{eq:MVDR output}) as
\begin{align}
\label{eq:MVDR-COPRA output 1}
y_{\text{BF}}[t] &= \frac{\av^H   \hat{\Cm}_{\yv\yv}^{-\frac{1}{2}}  \hat{\Cm}_{\yv\yv}^{-\frac{1}{2}} \yv }{\av^H \  \hat{\Cm}_{\yv\yv}^{-\frac{1}{2}}   \hat{\Cm}_{\yv\yv}^{-\frac{1}{2}} \av}
=
\frac{\bv^H \zv}{\bv^H \bv},
\end{align}
where $\bv \triangleq  \hat{\Cm}_{\yv\yv}^{-\frac{1}{2}} \av$ and $\zv \triangleq  \hat{\Cm}_{\yv\yv}^{-\frac{1}{2}} \yv$. These two relationships can be thought of as the inverses of the linear systems
\begin{align}
\label{eq:MVDR-COPRA b}
\av =  \hat{\Cm}_{\yv\yv}^{\frac{1}{2}} \bv,
\end{align}
and
\begin{align}
\label{eq:MVDR-COPRA z}
\yv =  \hat{\Cm}_{\yv\yv}^{\frac{1}{2}} \zv.
\end{align}
Since the matrix $ \hat{\Cm}_{\yv\yv}^{\frac{1}{2}}$ is ill-conditioned, direct inversion does not provide a viable solution. Therefore, we can apply a regularization method to obtain an estimate of $\bv$ and $\zv$ based on the linear models \eqref{eq:MVDR-COPRA b} and \eqref{eq:MVDR-COPRA z}. Given that $\av$ and $\yv$ are noisy, each of the latter two models is equivalent to the linear model in (\ref{eq:model}). In this paper, we propose using a regularization algorithm to estimate $\bv$ and $\zv$ based on the models \eqref{eq:MVDR-COPRA b} and \eqref{eq:MVDR-COPRA z}. The results are then substituted in \eqref{eq:MVDR-COPRA output 1}. Using \eqref{eq:RLS} for $\Am = \hat{\Cm}_{\yv\yv}^{\frac{1}{2}}$ and the \emph{eigenvalue decomposition} (EVD) $\hat{\Cm}_{\yv\yv} = \Um\Sigmam^2\Um^H$, the beamformer output using RLS will take the form
\begin{align}
\label{eq:MVDR-RLS output}
y_{\text{BF-RLS}} =\frac{  \av^H \Um \left(\Sigmam^2 + \gamma_{\text{b}} \Id \right)^{-1}  \left(\Sigmam^2 + \gamma_{\text{z}} \Id \right)^{-1} \Sigmam^2 \Um^H \yv } { \av^H \Um \left(\Sigmam^2 + \gamma_{\text{b}} \Id \right)^{-2} \Sigmam^2 \Um^H \av },
\end{align}
where $\gamma_{\text{b}}$ and $\gamma_{\text{z}}$ are the regularization parameters pertaining to the linear systems \eqref{eq:MVDR-COPRA b} and \eqref{eq:MVDR-COPRA z}, respectively. Equation~\eqref{eq:MVDR-RLS output} suggests that the weighting coefficients for the RLS approach are given by
\begin{align}
\label{eq:MVDR-RLS weight}
\wv_{\text{BF-RLS}} = \frac{  \av^H \Um \left(\Sigmam^2 + \gamma_{\text{b}} \Id \right)^{-1}  \left(\Sigmam^2 + \gamma_{\text{z}} \Id \right)^{-1} \Sigmam^2 \Um^H  } { \av^H \Um \left(\Sigmam^2 + \gamma_{\text{b}} \Id \right)^{-2} \Sigmam^2 \Um^H \av }.
\end{align}
Existing regularization methods (e.g., L-curve, GCV, quasi) can be used to find $\gamma_{\text{b}}$ and $\gamma_{\text{z}}$ in (\ref{eq:MVDR-RLS weight}). In the following section, we introduce a new regularization approach called MVDR constrained perturbation regularization approach (MVDR-COPRA) that is based on exploiting the eigenvalue structure of $\hat{\Cm}_{\yv\yv}^{\frac{1}{2}}$ in order to find the regularization parameters required in \eqref{eq:MVDR-RLS weight}. To this end, we replace $\Am$ in (\ref{eq:model}) by $\hat{\Cm}_{\yv\yv}^{\frac{1}{2}}$ to obtain the model
%
%
%that is based on exploering the eign values of the the matrix $\Cm_{\xv\xv}$
%Now, the question is how to find $\gamma_{\text{b}}$ and $\gamma_{\text{z}}$  in (\ref{eq:MVDR-RLS weight}). In the following section, we introduce a new approach called MVDR constrained perturbation regularization approach (MVDR-COPRA) to find the regularization parameters required in \eqref{eq:MVDR-RLS weight}. 
\begin{equation}
\label{eq:model2}
\rv = \hat{\Cm}_{\yv\yv}^{\frac{1}{2}} \xv + \vv.
\end{equation}

\begin{figure*}[t]
\setcounter{equation}{21}
\begingroup
    \fontsize{9.25pt}{9.25pt} 
\begin{equation}
\label{eq:eta min 2}
\lambda_{\text{o}}^{2} \left(\sigma_{\vv}^{2} \  \text{tr}\left(\left(\Sigmam^{2}+\gamma_{\text{o}}\Id\right)^{-2}\right)+\text{tr}\left( \left(\Sigmam^{2} + \gamma_{\text{o}} \Id \right)^{-2}\Sigmam^{2} \Um^{H} \Cm_{\xv\xv} \Um  \right)\right) = 
{\sigma_{\vv}^{2} \ \text{tr}\left(\Sigmam^{2}\left(\Sigmam^{2}+\gamma_{\text{o}}\Id\right)^{-2}\right) +\text{tr}\left( \Sigmam^{2} \left(\Sigmam^{2} + \gamma_{\text{o}} \Id \right)^{-2}\Sigmam^{2} \Um^{H} \Cm_{\xv\xv} \Um  \right)}
\end{equation}
\endgroup
\vspace{-10pt}
\begingroup
    \fontsize{9.2pt}{9.2pt} 
\begin{align}
\label{eq:eta min 3}
\lambda_{\text{o}}^{2}  &\approx \frac{
\sigma_{\vv}^{2} \text{tr}\left(\Sigmam_{1}^2 \left(\Sigmam_{1}^2 + \gamma_{\text{o}} \Id_{1} \right)^{-2} \right) + \text{tr} \left( \Sigmam_{1}^{2}\left(\Sigmam_{1}^2 + \gamma_{\text{o}} \Id_{1} \right)^{-2}\Sigmam_{1}^{2} \Um_{1}^{H} \Cm_{\xv\xv} \Um_{1} \right)}{\sigma_{\vv}^{2} \  \text{tr}\left(\left(\Sigmam_{1}^2 + \gamma_{\text{o}} \Id_{1} \right)^{-2} \right) + \frac{\sigma_{\vv}^{2} n_{2}}{\gamma_{\text{o}}^{2}} + 
\text{tr} \left(\left(\Sigmam_{1}^2 + \gamma_{\text{o}} \Id_{1}  \right)^{-2} \Sigmam_{1}^{2}  \Um_{1}^{H} \Cm_{\xv\xv} \Um_{1} \right) }
\end{align}
\endgroup
\vspace{-12pt}
%\hrulefill
\end{figure*}
%%%%%%%%%%%%%%%%%%%%%%%%%%%%%%%%%%%%%%%%%%%%%%%%%%%%%%%%%%%%%%%%%%%%%%%%%%%%%%%%%%%%%%%%%%%%%%%%%%%%%%%%%%%%%%%%%%%%%%%%%%%%%%%%%%%%%%%%%%%%%%%%%%%%%%%%%%%%%%%%%%%%%%%%%%%%%%%%%%%%%%%%%%%%%%%%%%%%%%%%%%%%%%%%%%%%%%%%
\section{The Proposed MVDR-COPRA }
\label{se:mse}
Due to the ill-conditionedness property of $\hat{\Cm}_{\yv\yv}^{\frac{1}{2}}$, we propose perturbing the model in (\ref{eq:model2}) as
\setcounter{equation}{12}
\begin{equation}
\label{eq:BDU equation}
\rv \approx \left(\hat{\Cm}_{\yv\yv}^{\frac{1}{2}}+ \Deltam \right)\xv + \vv,
\end{equation}
where $\Deltam \in \mathbb{C}^{m \times n}$ is an unknown perturbation matrix. We envision that this perturbation will alter the structure of the eigenvalues of $\hat{\Cm}_{\yv\yv}^{\frac{1}{2}}$ and improve its condition number. Hence, it stabilizes the solution of (\ref{eq:BDU equation}) compared to that of (\ref{eq:model2}) and improves the final result. However, adding this perturbation results in a degree of loss in model fidelity which may overweight the aforementioned benefits. Thus, we need to stay close to the original model. To achieve this, we bound $\Deltam$ by a positive number $\lambda$ (i.e., $||\Deltam||_{2} \leq \lambda$). This bound is unknown and has to be chosen judiciously. For now, we assume that $\lambda$ is known and we will revert to this issue later.

To obtain an estimate of $\xv$, we consider minimizing the worst-case residual function of (\ref{eq:BDU equation}) as
\begin{eqnarray}
\label{eq:worst-error}
&\underset{\hat{\xv}}{\operatorname{\min}} \  \underset{\Deltam}{\operatorname{\max}}  \ ||\rv - \left(\hat{\Cm}_{\yv\yv}^{\frac{1}{2}}+ \Deltam \right) \hat{\xv} ||_2 \nonumber \\
&\text{subject to} \,\, ||\Deltam||_2 \leq \lambda.
\end{eqnarray}
It can be shown that solving (\ref{eq:worst-error}) is equivalent to
\begin{equation}
\label{eq:costfunction}
\underset{\hat{\xv}}{\operatorname{\min}} \ \underbrace{||  \rv  -\hat{\Cm}_{\yv\yv}^{\frac{1}{2}} \hat{\xv}||_2 + \lambda \ ||\hat{\xv} ||_2}_{F\left(\hat{\xv}\right)}.
\end{equation}
Thus, the solution of (\ref{eq:worst-error}) is agnostic to the structure of $\Deltam$ and depends only on $\lambda$. The gradient of $F\left(\hat{\xv}\right)$ can be obtained as
\begingroup
    \fontsize{9.5pt}{9.5pt} 
\begin{align}
\label{eq:gradientI}
&\nabla_{\hat{\xv}} F\left(\hat{\xv}\right) \hspace{-7.5pt}
&=  \frac{1}{|| \rv- \hat{\Cm}_{\yv\yv}^{\frac{1}{2}} \hat{\xv}  ||_2} \Bigg[ \hat{\Cm}_{\yv\yv} \hat{\xv}  +  \frac{ \lambda   \ || \rv- \hat{\Cm}_{\yv\yv}^{\frac{1}{2}} \hat{\xv}  ||_2\ \hat{\xv} }{|| \hat{\xv} ||_2} 
&- \hat{\Cm}_{\yv\yv}^{\frac{1}{2}}\rv \Bigg]
\end{align}
\endgroup 
Now, define
\begin{equation}
\label{eq:sec1}
\gamma || \hat{\xv} ||_{2} =  \lambda \  || \rv -\hat{\Cm}_{\yv\yv}^{\frac{1}{2}} \hat{\xv} ||_2 .
\end{equation}
Substituting (\ref{eq:sec1}) in (\ref{eq:gradientI}) then solving $\nabla_{\hat{\xv}} F\left(\hat{\xv}\right) = 0 $ yields
\begin{equation}
\label{eq:RLS_BDU}
\hat{\xv} = \left( \hat{\Cm}_{\yv\yv} + \gamma\Id \right)^{-1}  \hat{\Cm}_{\yv\yv}^{\frac{1}{2}} \rv,
\end{equation}
which is the RLS estimator. From (\ref{eq:sec1}), we find that $\gamma$ is a function of $\lambda$ and the residual error of $\hat{\xv}$. Note that both $\lambda$ and $\hat{\xv}$ are unknowns. In the following, we show how to obtain $\gamma$ that corresponds to an optimal choice of $\lambda$. To this end, let us substitute for $\hat{\xv}$ from (\ref{eq:RLS_BDU}) in (\ref{eq:sec1}) and manipulate to obtain 
\begin{align}
% \begin{split}
\label{eq:secularEq1}
& \gamma^{2} \rv^{H}\hat{\Cm}_{\yv\yv}^{\frac{1}{2}} \left(\hat{\Cm}_{\yv\yv} + \gamma\Id\right)^{-2} \hat{\Cm}_{\yv\yv}^{\frac{1}{2}}\rv \nonumber\\
& - \lambda^{2} \left(\rv^{H} \rv - \rv^{H} \hat{\Cm}_{\yv\yv}^{\frac{1}{2}}\left(\hat{\Cm}_{\yv\yv} + \gamma\Id\right)^{-1}\hat{\Cm}_{\yv\yv}^{\frac{1}{2}}\rv   \right.\nonumber\\
& \left.- \gamma  \rv^{H} \hat{\Cm}_{\yv\yv}^{\frac{1}{2}}\left(\hat{\Cm}_{\yv\yv} + \gamma\Id\right)^{-2}\hat{\Cm}_{\yv\yv}^{\frac{1}{2}}\rv \right) = 0.
 %\end{split}
\end{align}
Next, we simplify (\ref{eq:secularEq1}) by applying the EVD of $\hat{\Cm}_{\yv\yv}$. Finally, we set $\gamma = \gamma_{\text{o}}$, $\lambda = \lambda_{\text{o}}$, and solve for $\lambda_{\text{o}}^{2}$ to obtain
\begin{equation}
\label{eq:eta min}
\lambda_{\text{o}}^{2} = \frac
{\text{tr}\left( \Sigmam^{2} \left(\Sigmam^{2} + \gamma_{\text{o}} \Id \right)^{-2} \Um^{H} \left(\rv \rv^{H}\right) \Um  \right)}
{\text{tr}\left( \left(\Sigmam^{2} + \gamma_{\text{o}} \Id \right)^{-2} \Um^{H} \left(\rv \rv^{H}\right) \Um  \right)},
\end{equation}
where $\text{tr}\left(.\right)$ denotes the trace operator. Now, let us think of $\lambda_{\text{o}}$ as an average value over many realizations of the observation vector $\rv$. Based on this, we can replace $\rv \rv^{H}$ by its expected value $\mathbb{E} \left(\rv \rv^{H} \right)$, which can be written using (\ref{eq:model2}) as
\begin{eqnarray}
\label{•eq:yy'}
\mathbb{E} \left(\rv \rv^{H} \right) =\Um \Sigmam \Um^{H}\Cm_{\xv\xv} \Um \Sigmam \Um^{H}+ \sigma_{\vv}^{2} \Id,
\end{eqnarray}
where $\Cm_{\xv\xv} \triangleq \mathbb{E}\left(\xv \xv^{H} \right)$ is the covarince matrix of $\xv$. Substituting (\ref{•eq:yy'}) in (\ref{eq:eta min}) and manipulating yields (\ref{eq:eta min 2}).

To step up, the eigenvalues of the ill-conditioned matrix $\hat{\Cm}_{\yv\yv}^{\frac{1}{2}}$ can be divided into two groups of \emph{significant} eigenvalues and nearly zero eigenvalues. Based on this, the matrix $\Sigmam$ can be divided into two diagonal sub-matrices, $\Sigmam_{1}$, which contains the significant $n_{1}$ diagonal entries, and $\Sigmam_{2}$, which contains the last $n_{2} = n_{\text{e}} - n_{1}$ trivial diagonal entries. This splitting idea is introduced as a general case from the special one where all the eigenvalues are significant (i.e., $n_{1}=n_{\text{e}}$ and $n_{2}=0$) and no truncation is required. The threshold for truncating the eigenvalues can be obtained as the mean of the eigenvalues multiplied by a certain constant $\rho \in \left(0, 1\right)$. Similarly, we can write $\Um = [ \Um_{1} \  \Um_{2} ]$, where $\Um_{1}\in \mathbb{C}^{n_{\text{e}}\times n_{1}}$, and $\Um_{2}\in \mathbb{C}^{n_{\text{e}}\times n_{2}}$. 

Due to the way we choose $n_{1}$ and $n_{2}$, we have $\lVert \Sigmam_{2} \rVert \approx 0$. As a result, by substituting the partitioning of $\Sigmam$ and $\Um$ in (\ref{eq:eta min 2}) and manipulating, we obtain (\ref{eq:eta min 3}).

The optimal perturbation bound $\lambda_{\text{o}}$ in (\ref{eq:eta min 3}) is a function of the unknowns $\gamma_{\text{o}}$, $\sigma_{\vv}^{2}$, and $\Cm_{\xv\xv}$. In the next subsection, we will use the mean-squared error (MSE) as criterion to eliminate this dependency and then to find an expression for obtaining $\gamma_{\text{o}}$.
\begin{figure*}[t]
\setcounter{equation}{28}
\begingroup
    \fontsize{9.25pt}{9.25pt} 
\begin{equation}
\label{eq:eta min 4}
\lambda_{\text{o}}^{2} \left(\text{tr}\left(\left(\Sigmam_{1}^2 + \gamma_{\text{o}} \Id_{1} \right)^{-2} \left(\Sigmam_{1}^2 + \frac{n_1 \sigma_{\vv}^{2}}{\text{tr}\left(\Cm_{\xv\xv}\right)} \Id_{1} \right)  \right) + \frac{n_2 n_1 \sigma_{\vv}^{2}
}{\gamma_{\text{o}}^{2}{\text{tr}\left(\Cm_{\xv\xv}\right)}} \right)\approx
{\text{tr}\left( \Sigmam_{1}^2 \left(\Sigmam_{1}^2 + \gamma_{\text{o}} \Id_{1} \right)^{-2} \left(\Sigmam_{1}^2 + \frac{n_1 \sigma_{\vv}^{2}}{\text{tr}\left(\Cm_{\xv\xv}\right)} \Id_{1} \right)  \right)}
\end{equation}
\endgroup
\begingroup
    \fontsize{9.25pt}{9.25pt} 
\begin{equation}
\label{eq:eta min 5}
\lambda_{\text{o}}^{2} \left(\text{tr}\left( \left(\Sigmam_{1}^2 + \gamma_{\text{o}} \Id_{1} \right)^{-2}  \left(\frac{n_{\text{e}}}{n_1}\Sigmam_{1}^2 + \gamma_{\text{o}} \Id_{1} \right) \right) +\frac{n_{2}}{\gamma_{\text{o}}}\right) \approx
\text{tr}\left( \Sigmam_{1}^2 \left(\Sigmam_{1}^2 + \gamma_{\text{o}} \Id_{1} \right)^{-2} \left(\frac{n_{\text{e}}}{n_1}\Sigmam_{1}^2 + \gamma_{\text{o}} \Id_{1} \right)  \right)
\end{equation}
\endgroup
\vspace{-7pt}
\begingroup
    \fontsize{9.25pt}{9.25pt} 
\begin{align}
\label{eq:secularEq2}
&\text{tr}\left( \Sigmam^2 \left(\Sigmam^2 + \gamma_{\text{o}} \Id \right)^{-2}\dv\dv^{H} \right)\text{tr}\left( \left(\Sigmam_{1}^2 + \gamma_{\text{o}} \Id_{1} \right)^{-2}\left(\beta\Sigmam_{1}^2 + \gamma_{\text{o}} \Id_{1}\right) \right) + \frac{n_{2}}{\gamma_{\text{o}}}\text{tr}\left( \Sigmam^2 \left(\Sigmam^2 + \gamma_{\text{o}} \Id \right)^{-2}\dv\dv^{H} \right) \nonumber\\
&-\text{tr}\left(\left(\Sigmam^2 + \gamma_{\text{o}} \Id \right)^{-2} \dv\dv^{H}\right)\text{tr}\left( \Sigmam_{1}^2 \left(\Sigmam_{1}^2 + \gamma_{\text{o}} \Id_{1} \right)^{-2} \left(\beta\Sigmam_{1}^2 + \gamma_{\text{o}} \Id_{1}\right) \right) = 0
\end{align}
\endgroup
\vspace{-15pt}
%\hrulefill
\end{figure*}

\subsection{Minimizing the MSE of the RLS}
Starting from the RLS in (\ref{eq:RLS_BDU}), we define the overall MSE as
\setcounter{equation}{23}
\begin{equation}
\label{eq:MSE1}
\text{MSE} =  \text{tr}\left\{ \mathbb{E}\left( (\hat{\xv} - \xv) (\hat{\xv} - \xv)^{H}  \right) \right\}.
\end{equation}
Substituting the EVD of $\hat{\Cm}_{\yv\yv}^{\frac{1}{2}}$ in (\ref{eq:RLS_BDU}) and plugging the result in (\ref{eq:MSE1}) we obtain
\begin{align}
\label{eq:MSE2}
&\text{MSE}
= \sigma_{\vv}^{2} \text{tr}\left(\Sigmam^{2} \left(\Sigmam^{2} + \gamma\Id \right)^{-2} \right) \nonumber \\
&+ \gamma^{2} \text{tr}\left( \left(\Sigmam^{2} + \gamma\Id \right)^{-2}\Um^{H}\Cm_{\xv\xv}\Um \right).
\end{align}
The MSE in (\ref{eq:MSE2}) is convex in $\gamma$, and its global minimzer (i.e., $\gamma_{\text{o}}$ ) can be obtained by solving
\begin{align}
\label{eq:MSE'}
&\frac{ \partial \left(\text{MSE}\right)}{\partial \  \gamma} 
= -2\sigma_{\vv}^{2} \text{tr}\left(\Sigmam^{2} \left(\Sigmam^{2} + \gamma\Id \right)^{-3} \right) \nonumber \\
&+
2 \gamma \ \text{tr}\left( \Sigmam^{2} \left(\Sigmam^{2} + \gamma\Id \right)^{-3}\Um^{H}\Cm_{\xv\xv}\Um \right) = 0.
\end{align}
However, this will not produce a closed-form expression for $\gamma_{\text{o}}$. To obtain a closed-form expression that is generaly feasible and also sub-optimal in some sense, we consider an average value for the second term in (\ref{eq:MSE'}) based on the inequalities in \cite{wang1986trace} (Equation (5)). As a result, (\ref{eq:MSE'}) can be approximated by
\begin{align}
\label{eq:MSE' approx}
&\frac{ \partial \left(\text{MSE}\right)}{\partial \  \gamma} 
\approx
-2 \sigma_{\vv}^{2} \text{tr}\left(\Sigmam^{2} \left(\Sigmam^{2} + \gamma\Id \right)^{-3} \right) \nonumber \\
&+ 
2 \gamma \frac{\text{tr}\left(\Cm_{\xv\xv} \right)}{n_{\text{e}}}\text{tr}\left(\Sigmam^{2} \left(\Sigmam^{2} + \gamma\Id \right)^{-3} \right) = 0.
\end{align}
Solving (\ref{eq:MSE' approx}) yields
\begin{equation}
\label{eq:gamma min approx}
\gamma_{\text{o}} \text{tr}\left(\Cm_{\xv\xv}\right) \approx n_{\text{e}} \sigma_{\vv}^{2}.
\end{equation}
Now, we return to (\ref{eq:eta min 3}) and apply the same inequalities from \cite{wang1986trace} to obtain (\ref{eq:eta min 4}). Based on \eqref{eq:gamma min approx}, Equation (\ref{eq:eta min 4}) allows us to eliminate the dependency of $\lambda_{\text{o}}$ in (\ref{eq:eta min 3}) on the unknowns $\sigma_{\vv}^{2}$ and $\Cm_{\xv\xv}$ by replacing $\frac{n_1 \sigma_{\vv}^{2}} {\text{tr}(\Cm_{\xv\xv})}$ by $\frac{n_1}{n_{\text{e}}}\gamma_{\text{o}}$ to obtain (\ref{eq:eta min 5}).
%replace This new relation allows us to
%
%Based on \eqref{eq:gamma min approx}, we can insert $\frac{n_1}{n_{\text{e}}}\gamma_{\text{o}}$ to replace $\frac{n_1 \sigma_{\vv}^{2}} {\text{tr}(\Cm_{\xv\xv})}$ in (\ref{eq:eta min 4}) and manipulate to obtain (\ref{eq:eta min 5}).

Equation (\ref{eq:eta min 5}) dictates the relationship between $\lambda_{\text{o}}$ and $\gamma_{\text{o}}$ that approximately minimizes the MSE. Substituting (\ref{eq:eta min 5}) back in (\ref{eq:eta min}) and manipulating, we obtain (\ref{eq:secularEq2}) where $\dv\triangleq \Um^{H}\rv$, and $\beta \triangleq \frac{n_{\text{e}}}{n_{1}}$. Equation~\eqref{eq:secularEq2} is solved to obtain the regularization parameter $\gamma_{\text{o}}$. In this paper, Newton's method \cite{zarowski2004introduction}, initialized using a \emph{small} positive value, is exclusively used to solve (\ref{eq:secularEq2}).

Now, we substitute $\rv=\av$ and $\rv=\yv$ in (\ref{eq:secularEq2}) to obtain $\gamma_{\text{b}}$ and $\gamma_{\text{z}}$, respectively, by solving the resulting equations independently using Newton's method. Finally, we substitute $\gamma_{\text{b}}$ and $\gamma_{\text{z}}$ in (\ref{eq:MVDR-RLS weight}) to obtain the weighting coefficients. 
%
%
%Now, to obtain $\gamma_{\text{b}}$ and $\gamma_{\text{z}}$ required in (\ref{eq:MVDR-RLS weight}), we substitute for $\rv=\av$ and $\rv=\yv$ in (\ref{eq:secularEq2}) respectively. Then, we solve the result equation by Newton's method. Finally, we substitute the result values in (\ref{eq:MVDR-RLS weight}) to obtain the weighting coefficients. 
\setcounter{equation}{31}
\section{Results}
The performance of the proposed MVDR-COPRA is evaluated in terms of the \emph{signal-to-interference-and-noise ratio} (SINR). The SINR is defined as \cite{li2003robust}
\begin{align}
\label{eq:SINR}
\text{SINR} = \frac{ \sigma_{\text{s}}^2 |\wv^H \av|^2} { \wv^H \Cm_{\left(i+n\right)}  \wv },
\end{align}
where $\sigma_{\text{s}}^2$ is the power of the signal of interest and $\Cm_{\left(i+n\right)}$ is the covariance matrix of the signal made up of the interference.
 
The simulation setup consist of a uniform linear array with 10 elements placed at half of the wavelength of the signal of interest and two interfering signals. At each simulation trial, the directions of arrival for the signal of interest and the interference are generated from a uniform distribution in the interval $[-90^{\text{o}}, 90^{\text{o}}]$. The steering vector $\av$ is calculated from the true direction of arrival of the signal of interest plus an error which is modeled to be uniformly distributed in the interval $[-5^{\text{o}}, 5^{\text{o}}]$. By randomizing both the angles and the error associated with them, we are attempting to avoid any situations that favors one algorithm or another. An SINR value is computed from $10^{3}$ trial. The proposed MVDR-COPRA is compared to a number of beamforming methods. Namely, we compare against the LCMV beamformer \cite{Frost1972, Takao1976}, the response vector optimization LCMV (RVO LCMV) beamformer \cite{Xu2015}, the MVDR based robust adaptive beamformer (RAB MVDR) \cite{Khabbazi2012}, and the robust adaptive beamformer based on semi-definite programming (RAB SDP) \cite{Yu2008}. We also plot the optimal performance when the true covariance matrix is known and the performance of the standard MVDR beamformer. In addition, we consider applying existing regularization methods to replace our method in solving the two linear least-squares problems involved in the proposed beamforming approach. However, we will present results only for the best performing method among them, which is the quasi-optimal\cite{morozov2012methods,bauer2008regularization}. 

\begin{figure}[h]
  \centerline{\includegraphics[width=  3.4in, height = 2.4in]{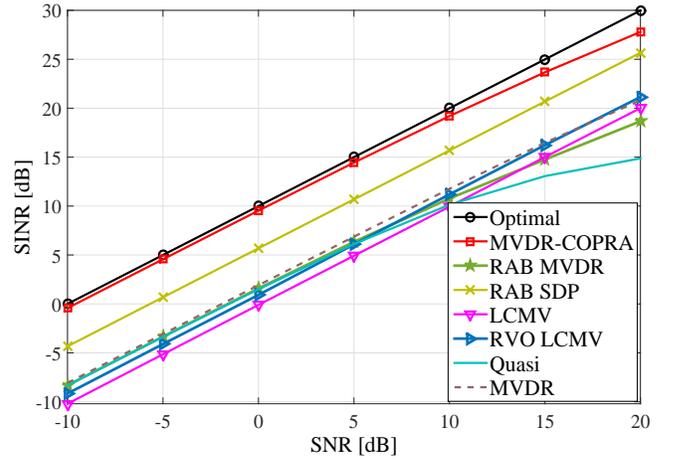}}  % Uncer3
\caption{Output SINR vs input SNR for $n_s=30$.}
\label{fig:uncer1}
\end{figure}

Fig.~\ref{fig:uncer1} plots the output SINR versus the input signal-to-noise ratio (SNR) for a number of snapshots $n_{s}=30$. It can be seen that the proposed MVDR-COPRA outperforms all the other methods by providing SINR that is very close to the optimal. Existing regularization methods are found to provide very poor performance. The best regularization method is the quasi optimal which offers very low SINR as Fig.~\ref{fig:uncer1} shows.
 
\begin{figure}[h]
  \centerline{\includegraphics[width=  3.4in, height = 2.4in]{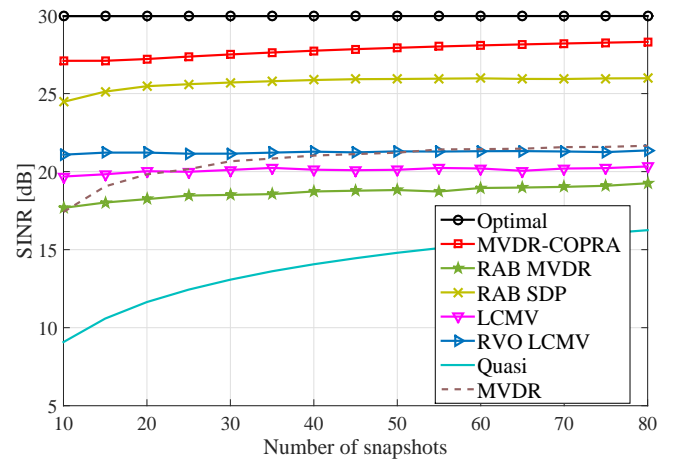}}  % Uncer3
\caption{Output SINR vs number of snapshots at SNR = 20 dB.}
\label{fig:uncer2}
\end{figure}

Fig.~\ref{fig:uncer2} plots the output SINR versus the number of snapshots $n_{s}$ for a fixed SNR of 20~dB. It is evident that the proposed MVDR-COPRA outperforms all the benchmark methods and stays close to the optimal performance. The closest to the MVDR-COPRA is the RAB SDP method. On the other hand, the best RLS method (quasi) offers very poor performance. 
\vspace{-8pt}
\section{Conclusion}
In this paper, we proposed a new approach for robust MVDR beamforming using regularized linear least-squares. Starting from the standard MVDR beamformer, the robust MVDR beamforming problem is converted to a pair of linear estimation problems with ill-conditioned matrices. We proposed a regularization method to solve these linear estimation problems. Simulation results demonstrate that the proposed approach outperforms a number of benchmark methods in terms of the signal-to-interference-and-noise ratio (SINR).
 
%\newpage 
%\clearpage

\bibliographystyle{IEEEbib}
\bibliography{refsGlobal}

\end{document}

%% file: macros.tex
\long\def\comment#1{}

\newfont{\bbb}{msbm10 scaled 700}

\newfont{\bb}{msbm10 scaled 1100}

\newcommand{\av}{{\bf a}}
\newcommand{\bv}{{\bf b}}

\newcommand{\dv}{{\bf d}}

\newcommand{\rv}{{\bf r}}

\newcommand{\wv}{{\bf w}}
\newcommand{\vv}{{\bf v}}
\newcommand{\xv}{{\bf x}}
\newcommand{\yv}{{\bf y}}
\newcommand{\zv}{{\bf z}}

\newcommand{\Am}{{\bf A}}

\newcommand{\Cm}{{\bf C}}

\newcommand{\Id}{{\bf I}}

\newcommand{\Um}{{\bf U}}

\newcommand{\Deltam}{\hbox{\boldmath$\Delta$}}
\newcommand{\Sigmam}{\hbox{\boldmath$\Sigma$}}